\documentclass[reprint,showpacs,showkeys,amsmath,amssymb,aps,nofootinbib]{revtex4-1}

\usepackage[american]{babel} 
\usepackage{graphicx}
\usepackage{amssymb} 
\usepackage{dcolumn}
\usepackage{bm}
\usepackage{color} 
\usepackage{dcolumn}
\usepackage{ulem} 

\def\lapproxeq{\lower .7ex\hbox{$\;\stackrel{\textstyle
<}{\sim}\;$}}
\def\gapproxeq{\lower .7ex\hbox{$\;\stackrel{\textstyle
>}{\sim}\;$}}

\begin{document}


\title{Coherent Radiation from Extensive Air Showers in the Ultra-High Frequency Band}

\author{Jaime Alvarez-Mu\~niz}
\author{Washington R. Carvalho Jr.}
\affiliation{
Depto. de F\'\i sica de Part\'\i culas
$\&$ Instituto Galego de F\'\i sica de Altas Enerx\'\i as,
Universidade de Santiago de Compostela, 15782 Santiago
de Compostela, Spain
}

\author{Andr\'es Romero-Wolf}
\affiliation{
Jet Propulsion Laboratory, California Institute of Technology, 4800 Oak Grove Drive, 
Pasadena, California 91109, USA
}

\author{Mat\'\i as Tueros}
\author{Enrique Zas}
\affiliation{
Depto. de F\'\i sica de Part\'\i culas
$\&$ Instituto Galego de F\'\i sica de Altas Enerx\'\i as,
Universidad de Santiago de Compostela, 15782 Santiago
de Compostela, Spain
}

\date{\today}
\begin{abstract}
{
Using detailed Monte Carlo simulations we have characterized the features of the radio emission
  of inclined air showers in the Ultra-High Frequency band (300 MHz -  3 GHz).
  The Fourier-spectrum of the radiation is
  shown to have a sizable intensity well into the GHz frequency range. The emission
  is mainly due to transverse currents induced by the geomagnetic field and 
  to the excess charge produced by the Askaryan effect.    
  At these frequencies only a significantly reduced volume of
  the shower around the axis contributes coherently to the
  signal observed on the ground. 
  The size of the coherently emitting volume depends on 
  frequency, shower geometry and observer position, 
  and is interpreted in terms of the relative time delays
  with respect to a plane shower front moving at the speed of light. 
  At ground level, the maximum emission at high frequencies
  is concentrated in an elliptical ring-like region around the
  intersection of a Cherenkov cone with its vertex at shower maximum and
  the ground.  The frequency spectrum of
  inclined showers when observed at positions that view shower maximum in 
  the Cherenkov direction, is shown to be in broad agreement with the pulses detected
  by the Antarctic Impulsive Transient Antenna (ANITA) experiment, 
  making the interpretation that they are due to Ultra-High Energy Cosmic Ray atmospheric showers 
  consistent with our simulations. 
  These results are also of great importance for experiments
  aiming to detect molecular bremsstrahlung radiation in the GHz
  range as they present an important background for its detection. }

\end{abstract}



\maketitle

\section{Introduction}

Radio pulses from ultra-high energy cosmic rays (UHECRs)
were first observed in the 1960s and 1970s in coincidence with air shower
arrays~\cite{jelley,allan} but severe difficulties were encountered to
extract sufficiently precise information about the showers themselves
using this technique alone~\cite{Fegan_review}. 
Current advances in fast signal digitization and increased computational power  
have opened up new possibilities to exploit the technique
and multiple efforts are being carried out to characterize these pulses in detail.
Experiments such as AERA~\cite{AERA} in the context of the Pierre Auger
Collaboration, LOPES~\cite{LOPES} in the context of KASCADE, and
LOFAR~\cite{LOFAR}, are regularly detecting pulses.
These experiments are typically operating in the $\sim$ 30-80 MHz range since the
emission is expected to be at least partially coherent at those frequencies. 
There are also efforts to detect GHz emission from EAS claimed to
originate from a different and yet unconfirmed mechanism: molecular bremsstrahlung~\cite{Gorham_molecular},
such as AMBER~\cite{Gorham_molecular}, MIDAS~\cite{MIDAS}, CHROME~\cite{CHROME} and
EASIER~\cite{microwave-auger}.

Recently several events in the 300-900~MHz frequency range were
serendipitously observed with the ANITA balloon-borne antenna array which was flown
over Antarctica.  These events have been claimed to
be consistent with emission from Extended Air Showers (EAS) with energies between 
$10^{18}$~--~$5\times10^{19}$~eV \cite{ANITA_UHECR}. 
Earlier claims of detection of UHECRs in the ultra-high frequency (UHF) band exist \cite{Fegan_Nature1,Spencer_Nature,Fegan_Nature2}.
Other experiments have also recently observed
GHz radiation associated with EAS~\cite{EASIER,CHROME} although  
at higher frequencies, typically above 3 GHz. 
These results motivate a careful study of the radio emission induced by
UHECR showers paying particular attention to the UHF band 
in the 300 MHz - 3 GHz range. 

The emission is due to both shower charges and currents:
an excess of electrons accumulates because there are no
positrons in the atmosphere, a mechanism 
predicted by Askaryan~\cite{Askaryan62}, and drift currents are induced
through the separation of positively and negatively charged particles 
in the magnetic field of the Earth~\cite{Kahn-Lerche,Allan_Nature,Falcke_radio_EAS,CODALEMA},
with the latter mechanism dominating the emission. 

When the induced currents and the net charge distributions move at a speed faster 
than the speed of light in the atmosphere, several features characteristic of Cherenkov emission
appear, including an enhancement of the high frequency components of the field \cite{ZHAireS,Scholten_arXiv}. 
Calculations of the emission from air showers have been performed using  
Monte Carlo simulations~\cite{ZHAireS,REAS3,SELFAS,COREAS}, 
and analytical and semi-analytical 
techniques~\cite{Scholten_MGMR,Seckel_ARENA2012}.

In this work we characterize the features of the UHF emission from EAS 
using ZHAireS \cite{ZHAireS}, a full Monte Carlo simulation 
of the air shower and its associated radio emission. 
The simulation is based on the AIRES shower code \cite{aires}, and the emission is calculated using 
a well established algorithm \cite{allan} obtained from first
principles~\cite{ZHS91,ZHS92,ARZ10} with which the electric field due to each
individual charged particle track produced in the shower simulation is calculated. 
In this process interference between the emission from
different tracks is carefully accounted for, including the  
variation of the refractive index $n$ with altitude. The algorithm does not assume any
specific emission mechanism and correctly reproduces the radiation induced by
the excess charge and the geomagnetic field \cite{ZHAireS}. 

\section{Characterization of radio emission in the UHF band}

\begin{figure}[htb]
\begin{center}
\scalebox{0.57}{
\includegraphics{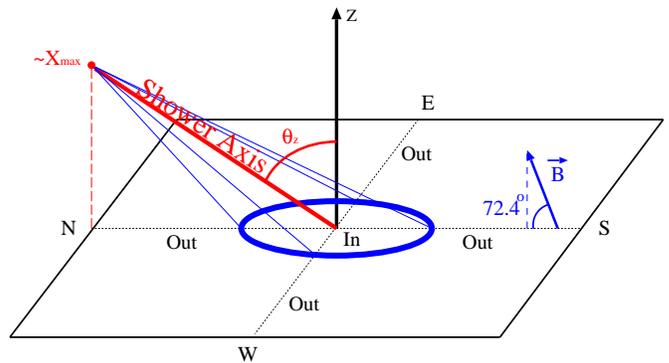}}
\vskip -0.5cm
\caption{Shower geometry:
  $\theta_z$ is the azimuthal angle of the shower, coming
  from the North. Antennas are placed along the E-W and N-S lines
  (dotted lines). The magnetic field
  $\vec{B}$ used in the simulations points towards North and has an
  inclination of -72.42$^\circ$. Also drawn is the Cherenkov cone centered
  at the depth of maximum shower development $X_{\rm max}$ and the ellipse of its intersection with the ground, 
  where the UHF signal is very close to its maximum value. This cone is present in any simulation with
  $n>1$ \cite{ZHAireS,Scholten_arXiv,Scholten_PRL}.}
\label{fig:geometry}
\end{center}
\end{figure}

For the emission in the UHF band, inclined showers are addressed in
more detail for two reasons: as they have the shower 
maximum high up in the atmosphere, the observation 
distance is usually quite far away and the conditions
for coherent emission from different stages of shower evolution 
are strengthened \cite{ZHS_thinned}. In addition it is inclined showers that have been 
observed in the GHz regime by ANITA. 
EAS have been simulated over Antarctica (altitude $\sim2800$ m above sea level
and atmospheric density profile for the South Pole)
with zenith angles between $\theta_z=50^\circ$ and $\theta_z=80^\circ$,
and radiation patterns at different positions on the
ground have been obtained. A 55 $\mu T$ magnetic field pointing towards North with an inclination of
-72.42$^\circ$ has been assumed. Antennas have been positioned on the ground 
along East-West (E-W) and North-South (N-S) lines intersecting
at the shower impact point, as shown in Fig.~\ref{fig:geometry}. 

Our simulations of inclined showers predict strong pulses in the nano-second scale
for antennas that view the depth of maximum shower development ($X_{\rm max}$) at angles very
close to the Cherenkov angle, i.e. antennas placed on an elliptical ``ring''
shown in Fig.~\ref{fig:geometry}, defined by the intersection 
of a Cherenkov cone centered at $X_{\rm max}$ and the ground. This is in agreement
with the results reported in~\cite{ZHAireS,Scholten_arXiv}. As the
observation point moves away from this region, 
to the inner or outer regions of the Cherenkov cone, there is a significant
broadening in time of the pulse, which is equivalent to a relative
lowering of the spectrum in the UHF band. 
\begin{figure}
\begin{center}
\scalebox{0.45}{
\includegraphics{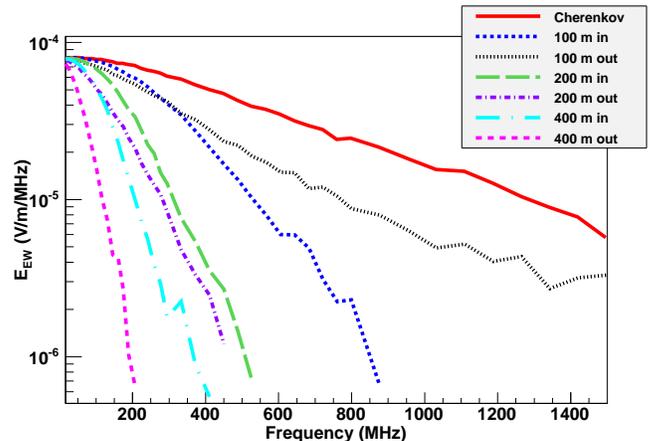}}
\vskip -0.5cm
\caption{Frequency spectra at ground 
for antennas at several distances along the W-E line that intercepts shower axis, inside and outside the Cherenkov cone 
for a $10^{19}$ eV proton shower with a zenith angle of $\theta_z=80^{\circ}$ coming from North. 
The antenna that sees $X_{\rm max}$ at the Cherenkov angle (solid red line) has a spectrum that extends well into the GHz frequency range.}
\label{fig:spectra}
\end{center}
\end{figure}
In Fig.~\ref{fig:spectra} we show the frequency spectra for antennas lying on
the ground along a W-E line that intercepts shower axis. 
The label ``Cherenkov'' refers to antennas that lay 
on the elliptical ring shown in Fig.~\ref{fig:geometry}, while numerical labels
refer to the distance in meters from the ring to the antenna, either towards
shower axis (in) or away from it (out). The spectra clearly become steeper as
the observation points get further away from the Cherenkov ring. 
Only antennas located very close to the ring contain a 
significant signal in the UHF band\footnote{Note there is an asymmetry between the ``in" and ``out" directions.}.

%

%
\begin{figure}[htb!]
\begin{center}
\scalebox{0.47}{
\includegraphics{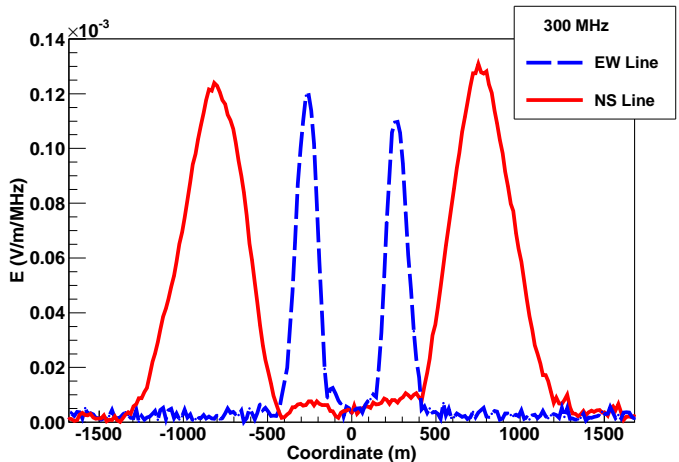}}
\vskip -0.5cm
\caption{Fourier component at 300 MHz as a function of distance to the shower core
  for a $10^{19}$ eV proton shower coming from the North with $\theta_z=70^\circ$. 
  The antennas were placed along the N-S and E-W lines that
  intersect at the shower core. Negative coordinates are South (West) of the
  core for antennas along the N-S (E-W) line.}
\label{fig:catedrales-ns-ew}
\end{center}
\end{figure}
\begin{figure}[htb!]
\begin{center}
\scalebox{0.47}{
\includegraphics{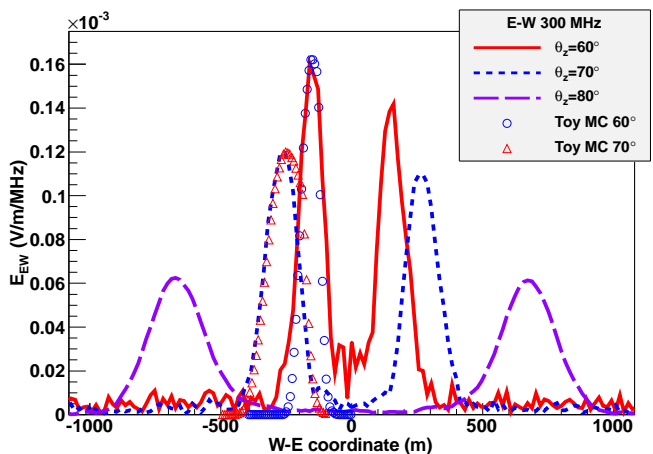}}
\vskip -0.5cm
\caption{Fourier component at 300 MHz as a function of distance to the core for a $10^{19}$ eV
  proton shower coming from the North with zenithal angles $\theta_z=60^\circ$, $70^\circ$ and
  $80^\circ$. The antennas were placed along an E-W line passing through the
  shower core (negative coordinates are West of the core). Also shown as points is the emission calculated using a simple
  one-dimensional toy model \cite{ZHAireS}, normalized to the fully simulated emission (see Sec.~\ref{sec:LDF} for details). 
The peaks West of the core are typically higher than to the East due to interference between the
  geomagnetic and Askaryan components~\cite{Scholten_MGMR,deVries,REAS3,ZHAireS}.}
\label{fig:catedrales-ew-multideg}
\end{center}
\end{figure}
In Fig.~\ref{fig:catedrales-ns-ew} we show the spectral component of the electric field 
at 300~MHz as a function 
of distance to the shower core for antennas along the S-N
and W-E axes. The shower comes from the North towards the South.  
The projection of the Cherenkov cone on ground makes 
an approximate ellipse with its major axis along the N-S
direction as expected.

\begin{figure}[htb!]
\begin{center}
\scalebox{0.47}{
\includegraphics{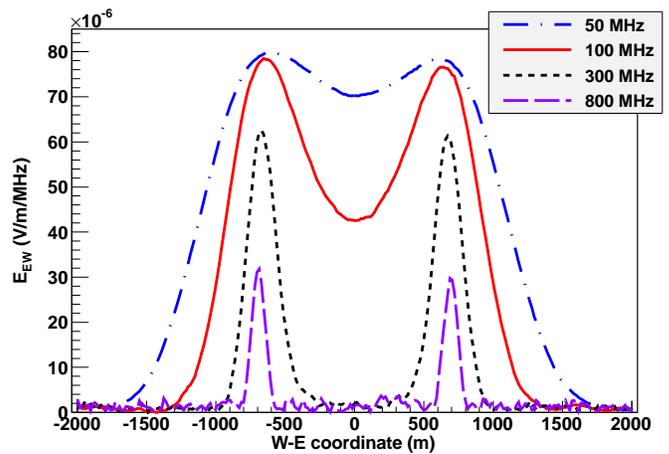}}
\vskip -0.5cm
\caption{Fourier components of the electric field 
at 50, 100, 300 and 800~MHz as a function of distance to the shower core for 
a $10^{19}$~eV proton shower coming from the North with zenith angle $80^\circ$. 
The antennas were placed along the E-W line passing through the impact point of the shower.}
\label{fig:catedrales-ew-multifreq}
\end{center}
\end{figure}

In Fig.~\ref{fig:catedrales-ew-multideg} we show the spectral components 
of the electric field for a $10^{19}$ eV proton shower with three different zenith angles. 
The scaling of
the major axis of the ring with $\sec\theta_z$ is illustrated. 
The axes and area of the ring increase as the zenith angle
rises because the shower maximum is more distant from the ground. This
increase dominates the drop induced because the opening angle of the Cherenkov 
cone (i.e. the Cherenkov angle) decreases as  
$n$ decreases with altitude\footnote{In the UHF band the Cherenkov angle at an altitude corresponding
to $X_{\rm max}$ decreases  
in the UHF band from $\sim1^\circ$ for a shower with zenith angle 50$^\circ$ to 
$\sim0.6^\circ$ for a shower with $\theta_z=80^\circ$.}.
In Fig.~\ref{fig:catedrales-ew-multifreq}  
the Fourier components of the field as a function of distance to the core are plotted for fixed zenith 
angle. As the frequency 
drops the angular width of the Cherenkov ring broadens (an effect already visible in
Fig.~\ref{fig:spectra}), and eventually it becomes broader than the Cherenkov angle itself 
making a ``plateau'' in the radial coordinate on ground. This can be mostly appreciated 
in the 50 MHz frequency line shown in Fig.~\ref{fig:catedrales-ew-multifreq}.  There is
evidence of this behaviour in the flattening of the lateral distribution of the signal 
close to the shower core in showers detected by the LOPES~\cite{LOPES_LDF} and LOFAR 
experiments \cite{LOFAR} as pointed out in \cite{Scholten_PRL,Scholten_arXiv}. 
%

\section{ANITA Observations}

The ANITA collaboration has recently reported observations
of UHECRs in the UHF band~\cite{ANITA_UHECR}. Most of these events
are ``reflected'' events, meaning that the radiation from the shower is
reflected on the Antarctic ice sheet towards the ANITA payload.
The ANITA observations exhibit power spectra (i.e. electric field squared) falling 
exponentially with frequency as $\exp(-\nu/\nu_0)$ between
300-900~MHz, with an average exponential constant $\nu_0 \sim 190$ MHz for reflected 
events\footnote{Note that $\nu_0 $ ($\nu_0/2$) 
is the constant of the exponential fall off of the electric field power (amplitude) spectrum.}. 
The results shown in Fig.~\ref{fig:spectra} for the electric field
amplitude, when squared, fit well to an exponential fall-off in the UHF band with an
exponential constant of $\nu_0 \sim 263$ MHz for an antenna located 
precisely in the Cherenkov ring, decreasing to $\nu_0 \sim 169$ MHz
(84 MHz) for the antenna 100 m away towards the outside (inside) of the Cherenkov ring. 
The interpretation that the radiation detected with the ANITA instrument 
was due to extensive air showers induced by UHECRs \cite{ANITA_UHECR}, 
is consistent with our simulations. The geometry must be 
such that the antenna is pointing in the direction of the 
reflected radiation emitted at the Cherenkov cone at shower maximum.

The ability of ANITA to observe events reflected 
from regions away from the Cherenkov ring on the ground is trigger limited.
The ANITA trigger requires coincidences between several frequency
bands registered by neighboring feeds as expected from impulsive signals. 
As a result the detector selects only those pulses with significant
power over multiple bands of width ranging from $130$ to $415$~MHz 
centered at frequencies 265, 435, 650, and 990 MHz~\cite{ANITA_instrument_paper}. 
A steeply falling cosmic-ray energy spectrum favors triggers on events with
flatter frequency spectra such as those expected near the Cherenkov ring.

Although we have only discussed the features of the spectrum on ground, 
we expect the reflection on the 
ice sheet as well as the larger distance from the shower to the ANITA payload 
when compared to the distance to ground, to reduce the strength of the spectrum 
but not to modify significantly the spectral features. 

\section{Interpretation: time delays}

In order to interpret the features of the frequency spectrum on 
the ground as obtained in the full Monte Carlo simulations,  
regions of the shower that emit {\it coherently} at a given
frequency have been studied. For this purpose, we have calculated for 
a given observer, the delays 
between the arrival time of the emission originated at different positions of the shower,  
with respect to the arrival of the earliest signal. For an observer in the Cherenkov 
ring, the earliest part of the signal originates at the shower axis, close to shower 
maximum. In order to contribute coherently at a
given frequency the magnitude of the delays must be below a fraction of  
the period $T$ of that frequency. For the purposes of discussion 
we fix that fraction to one half of the period.

In the limit of very large distances to the shower, the Cherenkov condition
guarantees that all points along the shower 
axis emit in phase for observation at the Cherenkov angle. However, 
this is not the case in a real situation with the observer at ground, since 
the shower as a whole is not observed in the far-field. In fact, as we move along 
the shower axis the angle of the line of sight to the observation 
point changes introducing time delays which induce destructive interference
between different stages in the longitudinal development of the shower \cite{ZHS_thinned}. 

It is convenient to separate the time delays into geometrical delays 
associated to the position from which the radiation originates, and intrinsic
time delays which are due to the fact that the shower front is lagging behind with
respect to a plane front that moves at the speed of light because of
both curvature and time spread. Geometrical delays can be
calculated analytically. The time delay of the shower, as 
obtained in simulations with AIRES and measured in experiments \cite{conicalshape}, is approximately 
proportional to the radius near shower axis with a constant 0.2 ns/m. 
In the calculation of the time delays we accounted for an index of refraction 
that varies with altitude in the same way as that in the full Monte Carlo simulation.

\begin{figure*}[htb!]
\begin{center}
{
\scalebox{0.4}{\includegraphics{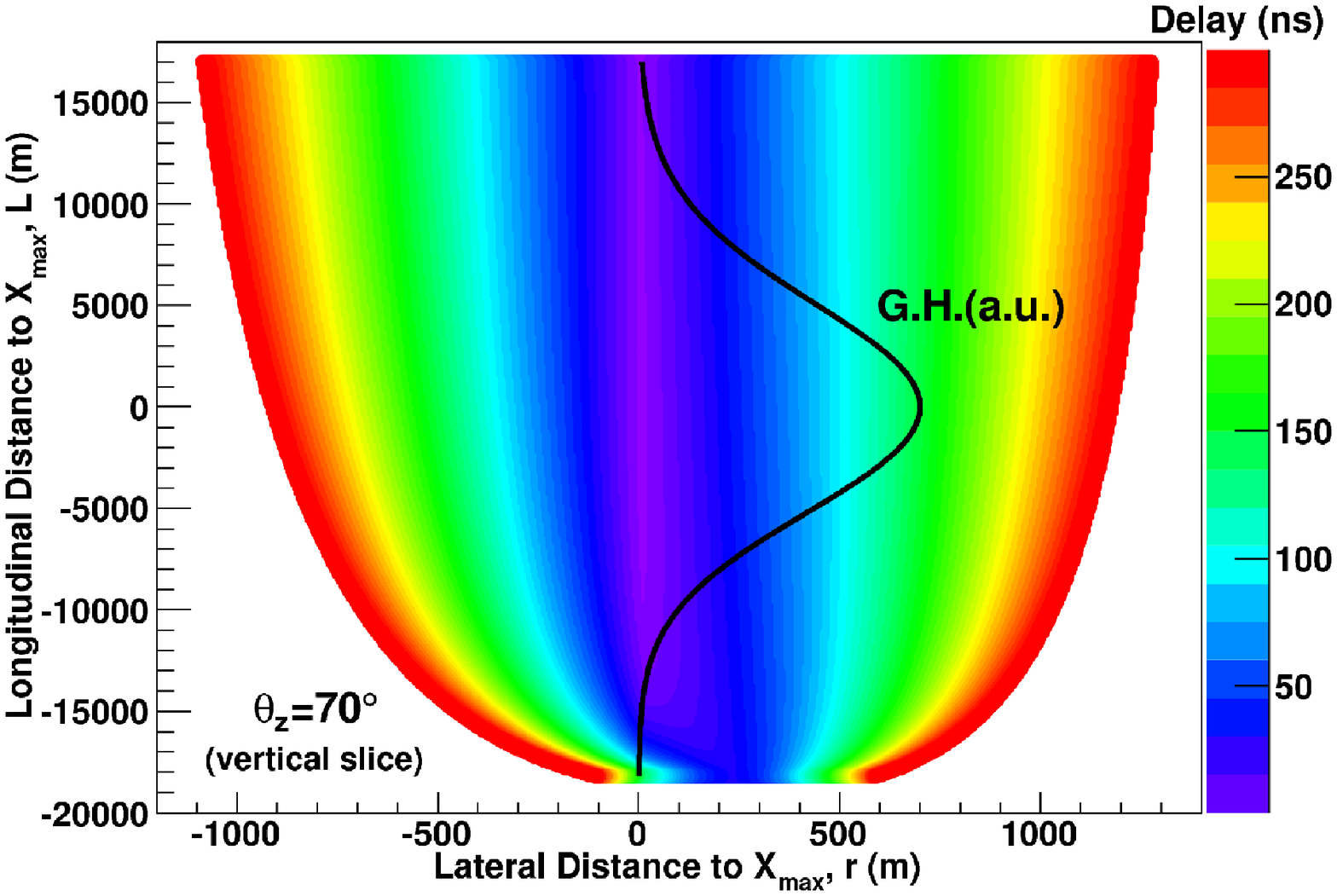}}
\scalebox{0.4}{\includegraphics{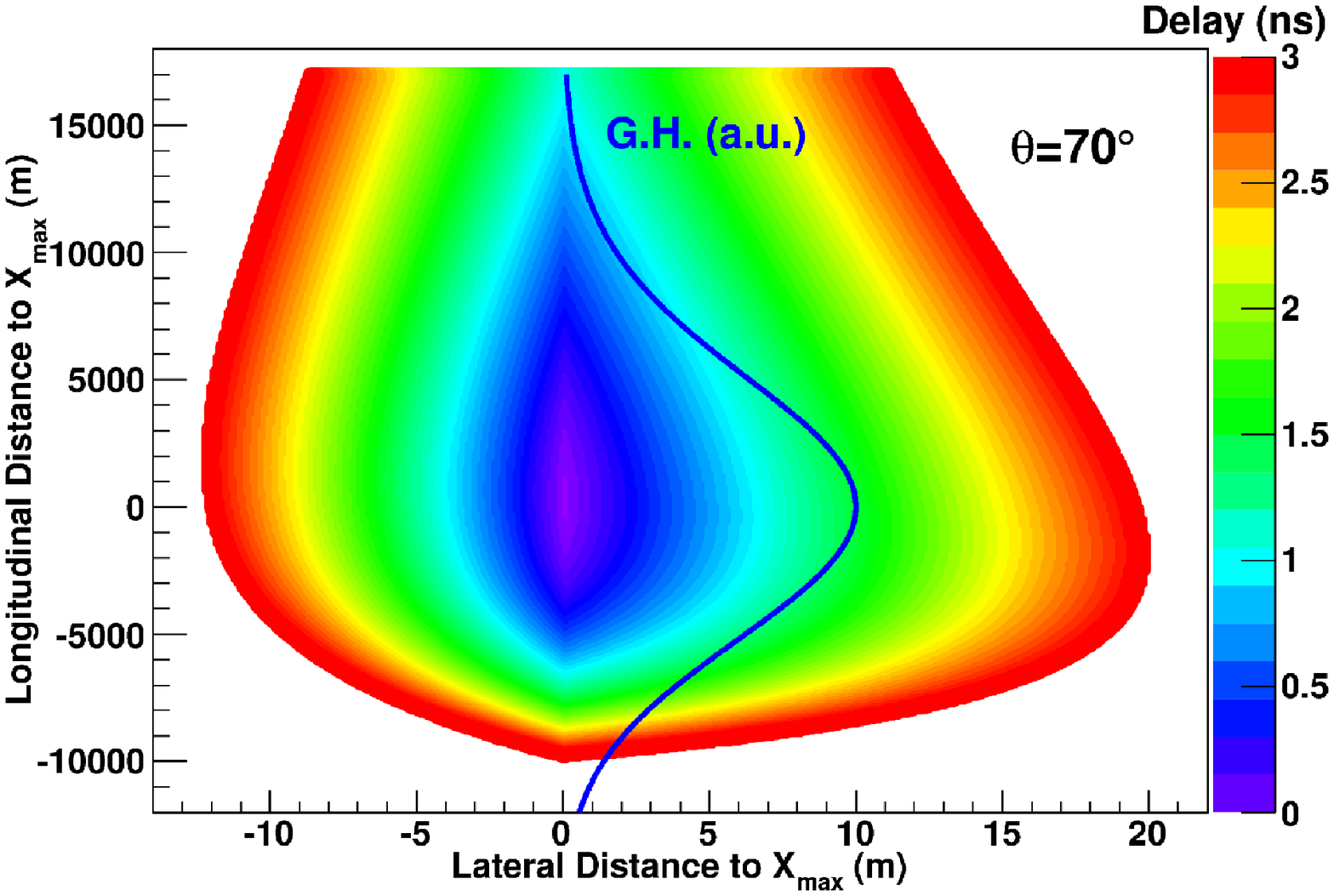}}
}
\caption{Time delays 
between the arrival time of the emission originated at different positions of the shower
- measured in the longitudinal $L$ and lateral $r$ directions -   
with respect to the arrival of the earliest signal.  
The delays were calculated for the geometry of a proton shower of energy $10^{19}~$eV with 
zenith angle $\theta_z=70^\circ$ simulated over Antarctica coming from the North, and for an  
observation point on the surface of the ice sheet at the South on the Cherenkov
ring. The coordinate $r$ measures the distance to the axis in the plane perpendicular to the  
shower incoming direction, and it runs from North (negative values of $r$) to South (positive values of $r$). 
At the observer position, the depth of maximum at the shower axis corresponding to the $r=0$ m, $L=0$ m 
position in the plot, is viewed at the Cherenkov angle. 
The longitudinal shower development extracted from the shower simulation 
is superimposed (solid line). Note that for this particular case the shower
starts at $\sim 17$ km measured along shower axis above $X_{\rm max}$.
Left panel: Delays for the region where the bulk of the shower is contained which are below $300$ ns. 
Right panel: Blow-up of the transverse region to illustrate the delays relevant for the spectrum 
in the UHF band. Note the different scales of the longitudinal and transverse axes in both panels.}
\label{TimeDelays}
\end{center}
\end{figure*}

The total delays as a function of the position in the shower ($r$ distance to shower axis, $L$ distance to $X_{\rm max}$)
from which the radio emission originates are displayed in Fig.~\ref{TimeDelays}
for a $10^{19}$ eV shower with $70^\circ$ of zenith angle.
The shower is coming from the North and the observer is located 
on ground at the South and at a position such that $X_{\rm max}$ 
is viewed at the Cherenkov angle (see Fig.~\ref{fig:geometry}). The coordinate $r$ is measured in the plane perpendicular
to the shower axis along the North-South direction. A similar picture is obtained in the 
East-West direction.  
Clearly for this geometry it is first the lateral distribution that is responsible
for the loss of coherence as the frequency increases. It can be appreciated
that for the bulk of the shower - contained within the Moli\'ere radius ($\sim 250-300$ m 
for such a shower at $X_{\rm max}$) - the time delays are below $\sim 100$ ns 
so that for frequencies well below 10 MHz the emission from all the shower is coherent.

The central region of the shower is separated in the right
panel of Fig.~\ref{TimeDelays} to indicate the delay structure in the region 
relevant for UHF emission. The coherent region can be seen to extend to several
km along the shower axis to either side of shower maximum. As points separated 
from the axis are considered, the intrinsic time delay of the shower front 
dominates for observers that see $X_{\rm max}$ close to the Cherenkov angle. 
This limits the coherent volume to lateral distances smaller than $r \sim 2.5~\nu/(1~{\rm GHz})$ m
with $\nu$ the frequency of observation. 
As a consequence, the region contributing  
in phase is a long and thin volume along shower axis.  

\begin{figure}
\begin{center}
\scalebox{0.45}{
\includegraphics{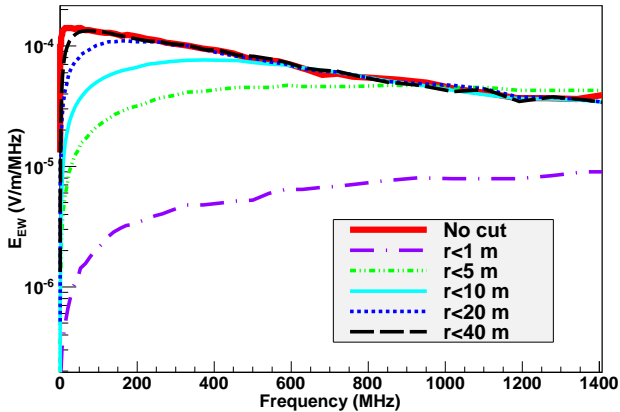}}
\vskip -0.5cm
\caption{Frequency spectra at ground 
for the antenna that views $X_{\rm max}$ at the Cherenkov angle  
for a $10^{19}$ eV proton shower with a zenith angle of $\theta_z=80^{\circ}$ coming from North
in Antarctica. 
The spectrum was obtained for the whole shower (``No cut") and for particles with 
distances to the core $r<r_{\rm cut}$ for values of $r_{\rm cut}=$1, 5, 10, 20 and 40 m. 
}
\label{fig:spectra_cut}
\end{center}
\end{figure}

This is confirmed in full Monte Carlo simulations in which 
we obtained the Fourier spectrum for a $10^{19}$ eV proton shower with $\theta_z=70^\circ$ 
and compared it to that obtained accounting only for particles occupying a fractional volume 
of the shower. For frequencies above 
$\sim 300$ MHz the spectrum of the whole shower is mainly due to particles with
$r < 10-20$ m as shown in Fig.~\ref{fig:spectra_cut}. 

This picture resembles the simple model described in~\cite{ZHAireS} which
neglects the lateral spread of the shower. In that model the 
high frequency signal was taken to be proportional to the value of the longitudinal
distribution at the point that ``views'' the antenna in the
Cherenkov direction. For inclined showers as is the focus of this work, 
this point must be replaced by an integral
over the longitudinal distance along shower axis 
from which the bulk of the radiation is produced coherently
at a given frequency in the UHF band. As the zenith angle decreases, the range of 
integration in the longitudinal dimension becomes
smaller and the picture approaches again the model in \cite{ZHAireS}.

\subsection{Lateral distribution of the field at ground level}
\label{sec:LDF}

For the observer viewing shower maximum at the Cherenkov angle,
the volume contributing coherently occupies a region of the shower
around $X_{\rm max}$ where the number of particles is largest.
This is illustrated in Fig.~\ref{TimeDelays} where the longitudinal 
shower profile is depicted on top of the spatial distribution of 
the delays. As a consequence, for this observer the electric field 
is largest.  If the observer moves along the 
radial direction in such a way that the  
shower maximum ceases to be in the Cherenkov direction, 
the volume contributing coherently moves upwards or downwards 
with respect to $X_{\rm max}$ into regions in the atmosphere where the number 
of charges contributing to the emission is typically smaller,
and the electric field is reduced.
This leads to the lateral distribution of the Fourier components of the
field shown in Figs.~\ref{fig:catedrales-ns-ew}, \ref{fig:catedrales-ew-multideg} and \ref{fig:catedrales-ew-multifreq}. 
Using the simple model in \cite{ZHAireS} accounting for the proper longitudinal region and 
the depth development of the number of charged particles in the cascade, 
the lateral distribution can be reproduced as shown in Fig.~\ref{fig:catedrales-ew-multideg}.
This indicates that the  
measurement of the spectral components in the UHF 
band at different distances to the impact point can be related to the 
longitudinal development of the shower.


\subsection{Frequency spectrum at ground level}

The approach of studying the time delays in the shower, allows us to interpret the frequency 
spectrum not only in the UHF band but at all frequencies, and all shower zenith angles 
including vertical showers.

\begin{figure*}[htb!]
\begin{center}
\scalebox{0.45}{
\includegraphics{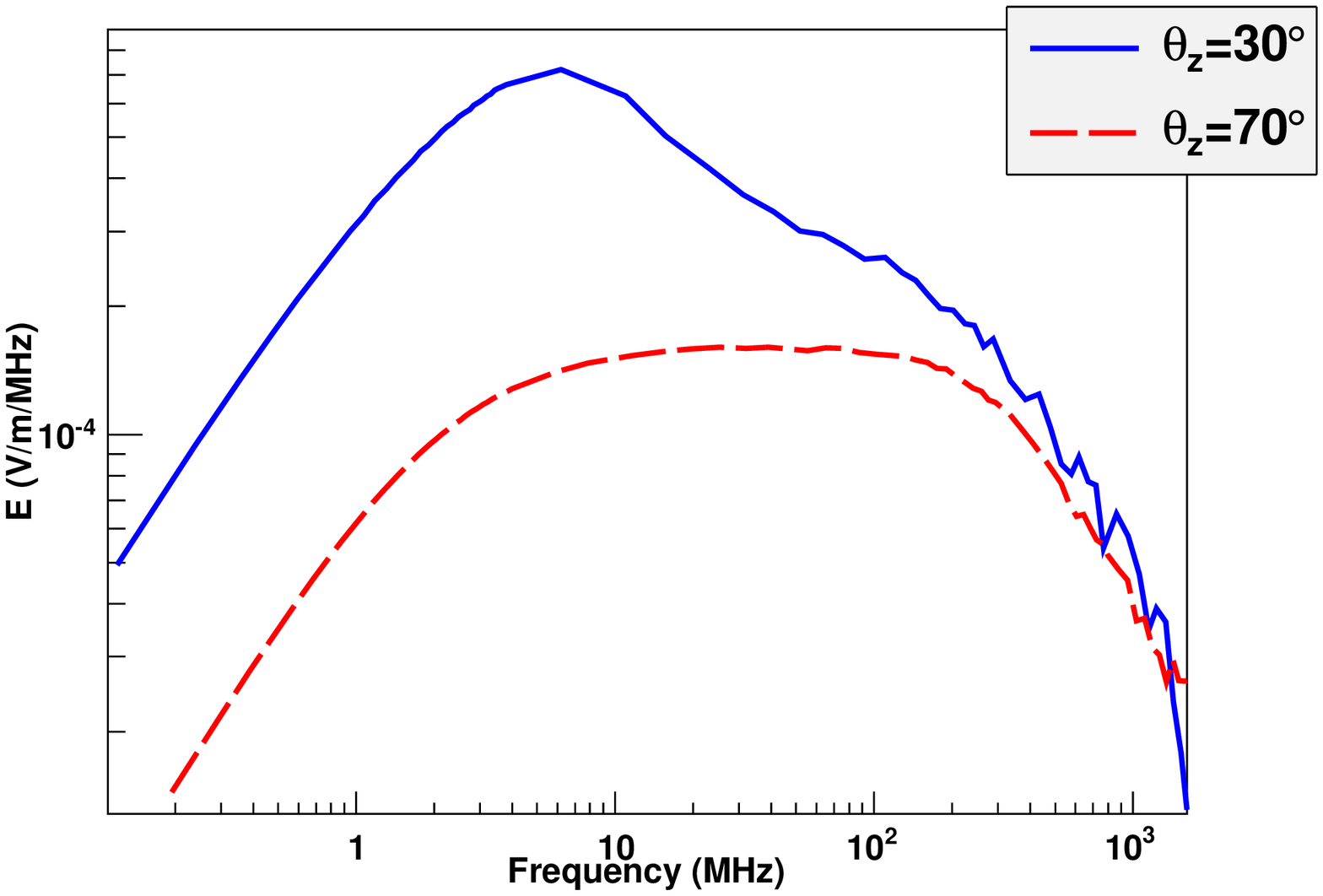} 
\includegraphics{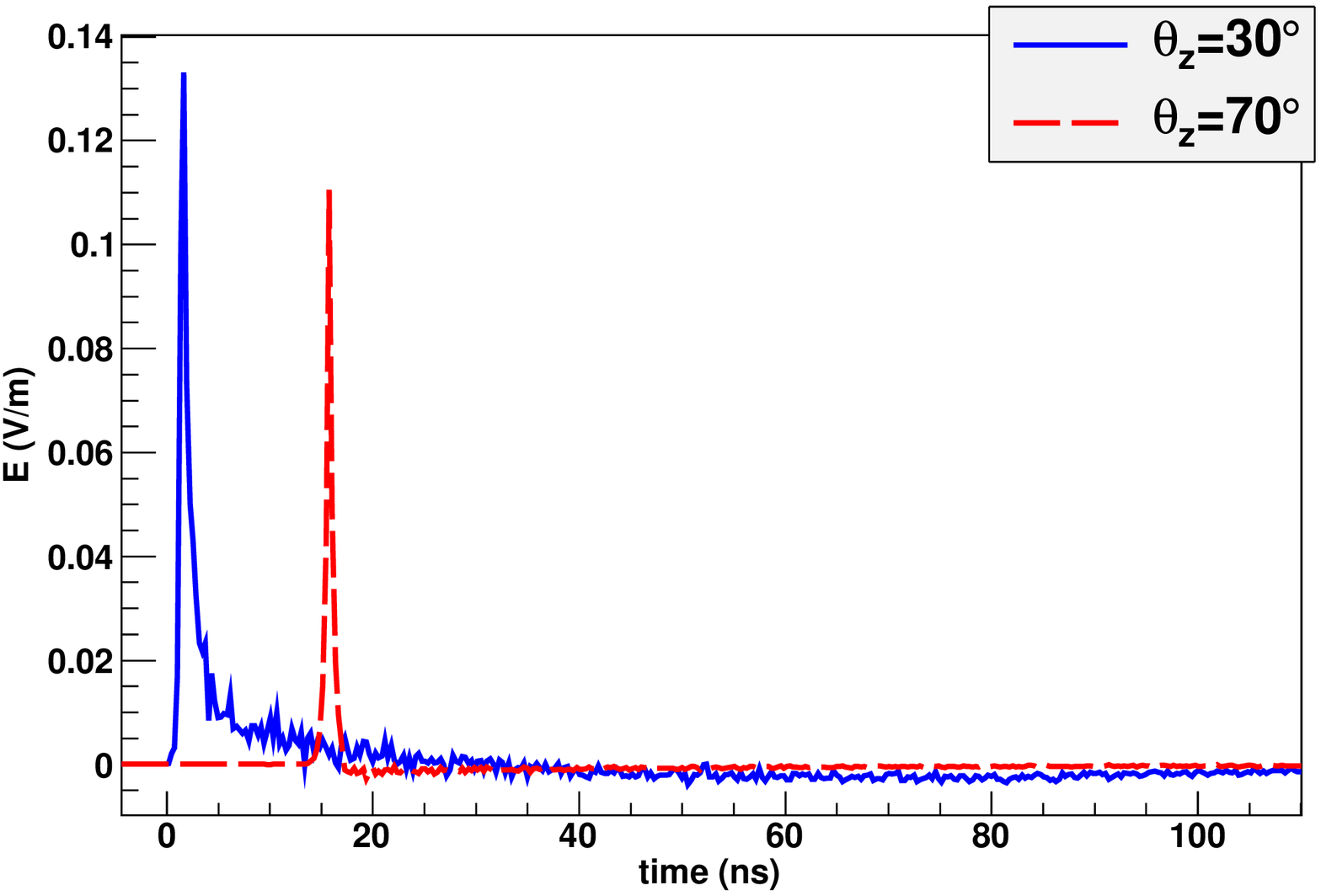}}
\caption{Left panel: frequency spectra at ground for 
a $\theta_z=30^\circ$ (top line) and inclined 
$\theta_z=70^\circ$ (bottom line) proton shower of energy 
$10^{19}$ eV in Antarctica. In both cases the observer
is viewing shower maximum at the Cherenkov angle.
Right panel: pulses in the time domain corresponding
to the frequency spectra shown in the left panel.
\label{fig:spectra2}}
\end{center}
\end{figure*}

The spectrum at ground from frequencies 
below 1 MHz up to GHz for a $10^{19}$ eV proton induced shower with $\theta_z=70^\circ$,  
and for an observer viewing shower maximum at the Cherenkov angle is shown in Fig.~\ref{fig:spectra2}.
Qualitatively, at very low frequencies the delays for any region in the shower are smaller 
than half the period of the corresponding frequency, and the coherent volume spans 
the whole shower. There is no loss in coherence from the volume of the shower contributing most to the emission as the frequency increases 
because it is constant with frequency. Since the electric field contributed by each particle track is proportional to frequency \cite{ZHS92}, 
the total field increases in the same fashion  for frequencies below $\sim 1-2$ MHz as can be seen in Fig.~\ref{fig:spectra2}.
As the frequency increases, the coherent volume starts to shrink but only in the lateral dimension. 
In the longitudinal direction and close to the shower axis, the delays are still small compared 
to half the period of the frequency because the observer is close to the Cherenkov angle for the bulk 
of the inclined shower. This can be seen in the left panel of Fig.~\ref{TimeDelays}. At $X_{\rm max}$ 
and far from the shower axis the delays are $\sim 300$ ns. This translates into a frequency $\sim 2$ MHz at
which the coherent volume starts to shrink laterally. The slope of the spectrum begins changing at that point 
due to the decrease in the number of particles contributing coherently. 
The change of slope is mild around 2 MHz because only particles outside the Moliere radius of the shower stop 
contributing coherently, and those constitute a small fraction of the shower, typically $< 20\%$. 
At some frequency the lateral dimension of the shrinking volume of the coherent region reaches the Moli\'ere radius 
($r_{\rm M}\sim 250-300$ m at $X_{\rm max}$ for a $70^\circ$ shower). 
Since the typical delay at the Moliere radius for a $70^\circ$ shower around $X_{\rm max}$ is $\sim 50$ ns, this occurs at 
a frequency of $\sim 10$ MHz. For $r<r_{\rm M}$ the lateral distribution is flatter than when $r>r_{\rm M}$,  
and the decrease in the number of particles contributing coherently is more important in relative terms as 
the frequency increases. As a consequence the spectrum flattens. 
This behavior is clearly seen in the spectrum obtained in full Monte Carlo simulations shown in Fig~\ref{fig:spectra2}. 
Near the lower edge of the UHF band, at $\sim 300$ MHz, only the volume inside which the delays are smaller than $\sim 1.5$ ns 
contributes coherently. This volume is depicted in the right panel of Fig.~\ref{TimeDelays} and it is of order 10 m lateral width.
As the frequency increases above $\sim 300$ MHz, the coherent volume shrinks in both the lateral and longitudinal dimensions, 
with the consequence that the spectrum starts dropping exponentially with frequency,  
as obtained with our full Monte Carlo simulations shown in Figs.~\ref{fig:spectra} and \ref{fig:spectra2}.

In essence, this is very similar for more vertical showers. 
The main difference is that the drop in the spectrum begins at a lower
frequency as can be seen in Fig.~\ref{fig:spectra2} where we show the spectrum 
for a $\theta_z=30^\circ$ proton shower of energy $10^{19}$ eV.
The reason for this is that in a more vertical shower the geometrical delays 
due to the longitudinal shower dimensions are typically larger than in the inclined case 
over the whole region in space occupied by the shower, and they start to become of the order
of the lateral and intrinsic delays at a smaller frequency.
As a result the time structure of the pulse has a broader
time scale for vertical showers and the high frequency components, although present, 
appear as a small modulation of the pulses. 

%

\section{Summary and outlook}

The radio signal emitted from UHECR air showers 
has been shown to have a sizable intensity well into the 
GHz frequency range. In relative terms
the UHF components are more important for inclined showers
than for vertical ones. 
When projecting on the ground, the peak of the UHF signal is located at 
a narrow elliptical ring defined by the intersection of the Cherenkov 
cone at shower maximum and the ground. The width of the ring becomes
narrower as the frequency is increased. The lateral distance to shower axis 
at which the UHF signal is recieved at highest amplitude depends on geometry, 
increasing with zenith angle, and reaching hundreds of meters for showers
of $70^\circ$. 

We have shown that the lateral distribution for the low frequency components 
of the radio pulses, typically below $\sim$100 MHz, displays a rather 
flat behavior with distance to the shower core until a position is reached at 
which shower maximum is viewed at the Cherenkov angle \cite{ZHAireS,Scholten_PRL}. 
This behavior is consistent with observations made at LOPES~\cite{LOPES_LDF} and 
LOFAR \cite{LOFAR} and should be taken into account when trying 
to obtain the shower geometry and properties from the radio pulse
measurements \cite{AERA}. 

The frequency spectrum of the pulses received in the
Cherenkov cone has been shown to extend well into the GHz regime and to
fall with an exponential slope in the UHF band, with a constant that 
is largest at the Cherenkov angle and drops as the observer views shower 
maximum away from the Cherenkov angle. Predictions for short radio pulses 
at ns scales follow from the spectra obtained for inclined showers. 

In the case of more vertical showers, although the UHF
components exist, the pulses are dominated by the lower frequency
components and thus give typically a much broader pulse in time. 

Our results indicate that by looking at the pulses at
different distances from the shower core with a dense array it could
be in principle possible to obtain information about the depth development
of the shower \cite{Scholten_arXiv}, but the detectors mut be positioned 
at fairly close distances because the Cherenkov cone is typically very narrow. 
These results provide an important background for experiments that are
trying to measure molecular bremsstrahlung radiation from extensive
air showers, and should be carefully considered in order to properly
interpret measurements made in the UHF band. It is worth noting that 
molecular bremsstrahlung is thought to be isotropic \cite{Gorham_molecular}, 
while the geomagnetic radiation
studied in this paper is highly collimated in the direction of the 
Cherenkov cone.  

The power spectra obtained in the Cherenkov
direction are consistent with the spectra measured at ANITA and
attributed to radio pulses from UHECR \cite{ANITA_UHECR}. The results of the simulations
made in this work further support this hypothesis. 
An event by event comparison of ANITA pulses and realistic predictions,
followed by a simulation of the experiment would further strengthen this claim.
The third flight of ANITA, scheduled for 2013-2014, is expected to
detect hundreds of cosmic rays~\cite{ANITA_UHECR}, 
and could provide a highly significant test of the expected UHECR
pulse properties. The EVA (Exa-Volt Antenna) project  
\cite{EVA} could increase the event rate by a factor of 10. 

\section{Acknowledgments}

J.A-M, W.R.C., M.T. and  E.Z. thank Xunta de Galicia (INCITE09 206 336 PR) and 
Conseller\'\i a de Educaci\'on (Grupos de Referencia Competitivos -- 
Consolider Xunta de Galicia 2006/51); Ministerio de  Educaci´on, Cultura y Deporte 
(FPA 2010-18410 and Consolider CPAN - Ingenio 2010); ASPERA (PRI-PIMASP-2011-1154) and Feder Funds, Spain.
We thank CESGA (Centro de SuperComputaci\'on de Galicia) for computing resources.
Part of this research was carried out at the Jet
Propulsion Laboratory, California Institute of Technology, under a contract with the National 
Aeronautics and Space Administration.


\end{document}